\documentclass[aps,superscriptaddress,amsmath,amssymb,twocolumn,showpacs,floatfix,english,noshowpacs]{revtex4}

\usepackage{url}
\usepackage{bm,bbm}
\usepackage{graphicx}
\usepackage{color}
\usepackage[colorlinks=true, urlcolor=blue, linkcolor=blue, citecolor=blue, pdftex]{hyperref}
\usepackage[english]{babel}

\usepackage{soul}
\usepackage{blindtext}

\newcommand{\dagga}{{\phantom{\dagger}}}

\begin{document}

\title{Gapless spin liquid and valence-bond solid in the $J_1-J_2$ Heisenberg model on the square lattice: insights from singlet and triplet
excitations}

\author{Francesco Ferrari}
%\email[]{ferrari@itp.uni-frankfurt.de}
\affiliation{Institute for Theoretical Physics, Goethe University Frankfurt, Max-von-Laue-Stra{\ss}e 1, D-60438 Frankfurt a.M., Germany}
\author{Federico Becca}
\affiliation{Dipartimento di Fisica, Universit\`a di Trieste, Strada Costiera 11, I-34151 Trieste, Italy}

\date{\today}

\begin{abstract}
The spin-$1/2$ $J_1-J_2$ Heisenberg model on the square lattice represents one of the simplest examples in which the effets of magnetic
interactions may suppress magnetic order, eventually leading to a pure quantum phase with no local order parameters. This model has been
extensively studied in the last three decades, with conflicting results. Here, by using Gutzwiller-projected wave functions and recently
developed methods to assess the low-energy spectrum, we show the existence of a level crossing between the lowest-energy triplet and 
singlet excitations for $J_2/J_1 \approx 0.54$. This fact supports the existence of a phase transition between a gapless spin liquid 
(which is stable for $0.48 \lesssim J_2/J_1 \lesssim 0.54$) and a valence-bond solid (for $0.54 \lesssim J_2/J_1 \lesssim 0.6$), even 
though no clear sign of dimer order is visible in the correlations functions. These results, which confirm recent density-matrix
renormalization calculations on cylindrical clusters [L. Wang and A.W. Sandvik, Phys. Rev. Lett. {\bf 121}, 107202 (2018)] reconcile 
the contraddicting results obtained within different approaches over the years.
\end{abstract}

\maketitle

\emph{Introduction-}
Since the beginning of the field of higly-frustrated magnetism, the spin-$1/2$ Heisenberg model, defined on the square lattice with both nearest- 
($J_1$) and next-nearest-neighbor ($J_2$) super-exchange couplings~\cite{chandra1988}, has played a pivotal role in the search and characterization 
of unconventional states of matter, in which correlation effects cannot be neglected~\cite{balents2010}. In the $J_1-J_2$ model, the combination 
of quantum and geometrical frustrations may lead to a ground state that has no local order parameters (such as a finite magnetization) and hosts 
elementary excitations with fractional quantum numbers. While the limiting case with $J_2=0$ (and equivalently $J_1=0$) can be assessed by 
Monte Carlo techniques and shows a conventional ground state with finite magnetization~\cite{reger1988,sandvik1997,calandra1998}, in presence of 
a finite frustrating $J_2$ term, there are no approaches that can give unbiased results on large lattices. Many different numerical techniques 
have been used to understand the nature of the ground state in the vicinity of $J_2/J_1=0.5$, where the highest level of frustration is expected 
(in the classical limit, where spins are tretaed as vectors of fixed length, this point presents a large degeneracy in the lowest-energy manifold). 

In early studies, some evidence for a valence-bond solid was obtained, from analytical approximations~\cite{read1989} and numerical calculations
on small clusters~\cite{figueirido1990,schulz1996}. Later, this scenario has been confirmed by series expansion techniques~\cite{singh1999}.
On the other hand, subsequent Monte Carlo calculations with Gutzwiller-projected fermionic wave functions hinted the possibility that the ground 
state of the system may be a pure quantum spin liquid, with no dimer order~\cite{capriotti2001}. More recently, several works addressed the 
question of the ultimate nature of the ground-state wave function in the highly-frustrated region $J_2/J_1 \approx 0.5$, with contraddicting
outcomes, either for a valence-bond solid~\cite{sushkov2001,mambrini2006,richter2010,gong2014,haghshenas2018} or 
a spin-liquid~\cite{jiang2012,wang2013,hu2013,poilblanc2017,liu2018,hering2019}. In this respect, recent developments in the numerical 
optimization of wave function with many parameters, such as tensor- or neural-network states, open promising routes to further investigate
this delicate issue~\cite{liao2019,choo2019,ferrari2019b,szabo2020}. 

Remarkably, in a recent paper on the subject, Wang and Sandvik highlited the existence of a level crossing in the low-energy spectrum, suggesting 
the possibility that the non-magnetic region of the model may consist of two phases: a gapless spin liquid, which develops continuously after the N\'eel state, 
and a valence-bond solid, which is stabilized for larger values of the frustrating ratio~\cite{wang2018}. In particular, by applying the 
density-matrix renormalization group technique to $2L \times L$ cylinders, they showed the existence of a level crossing between the lowest-energy 
triplet and singlet excitations for $J_2/J_1 \approx 0.52$. Since the N\'eel order is expected to disappear for $J_2/J_1 \gtrsim 0.46$, this 
feature has been associated to the transition between a gapless spin liquid and a valence-bond solid, likewise, in the one-dimensional $J_1-J_2$ 
model, a similar level crossing at $J_2/J_1 \approx 0.24$ marks the transition between the gapless (critical) phase and the dimerized 
one~\cite{okamoto1992,castilla1995}.

Here, we re-examine the issue of the level crossing between the lowest-energy triplet and singlet within variational wave functions constructed 
by using Abrikosov fermions, subject to the Gutzwiller projection that enables us to work in the correct Hilbert space of the spin 
model~\cite{wen2002}. Besides the standard calculations for the ground-state wave function, we apply the variational method proposed by Li and 
Yang~\cite{li2010}, and recently developed by us~\cite{ferrari2018b,ferrari2019a}, to construct both low-energy triplet and singlet states. 
The main motivation is to give an independent validation of the claim of Wang and Sandvik, considering, instead of $2L \times L$ cylindical 
geometries (with open boundary conditions along the $x$ direction), $L \times L$ square clusters with periodic boundary conditions, thus retaining 
all the symmetries of the infinite lattice. This fact allows us to identify the momenta of the excitations, which are $q=(\pi,\pi)$ 
for the triplet and $q=(\pi,0)$ [or $q=(0,\pi)$] for the singlet. In our previous work on the dynamical structure factor~\cite{ferrari2018b}, 
evidence for a gapless triplet excitations up to $J_2/J_1 \approx 0.55$ has been reported, supporting the existence of a spin liquid with 
Dirac-like spinon excitations (and ${\cal Z}_2$ gauge excitations)~\cite{hu2013}. Then, the main outcome of the present work is to confirm the 
presence of a singlet-triplet level crossing, which appears for $J_2/J_1 \approx 0.54$, in excellent agreement with what has been obtained by 
Wang and Sandvik~\cite{wang2018}. However, within the clusters that are presently available (i.e., $L$ up to $20$), a precise size-scaling 
extrapolation of the gaps is extremely difficult, not excluding a vanishing value in the thermodynamic limit in the entire non-magnetic phase. 
In this regard, dimer-dimer correlations of the ground state are compatible with a power-law decay both before and after the level crossing, 
suggesting a very large correlation length and a tiny dimer-order parameter. Therefore, it turns out that, as for the one-dimensional $J_1-J_2$ 
model, the phase transition is much more easily detected by looking at the level crossing, instead of looking at ground-state properties or at 
the triplet gap alone. Still, we mention that our calculations (as well as the ones performed by Wang and Sandvik) cannot exclude that the level 
crossing, in two dimensions, instead of signaling the onset of a valence-bond solid, could mark an unconventional transition between two 
paramagnetic phases, both supporting gapless singlet and triplet excitations.

%%%%%%%%%%%%%%%%%%%%%%%%%%%%%%%%%%%%%%%%%%%%%%%%%%%%%%%%%%%%%%%%%%%%%%%%%%%%%%%%%%%%%%%%%%%%%%%%%%%%%%%%%%%%%%
\begin{figure}
\includegraphics[width=0.9\columnwidth]{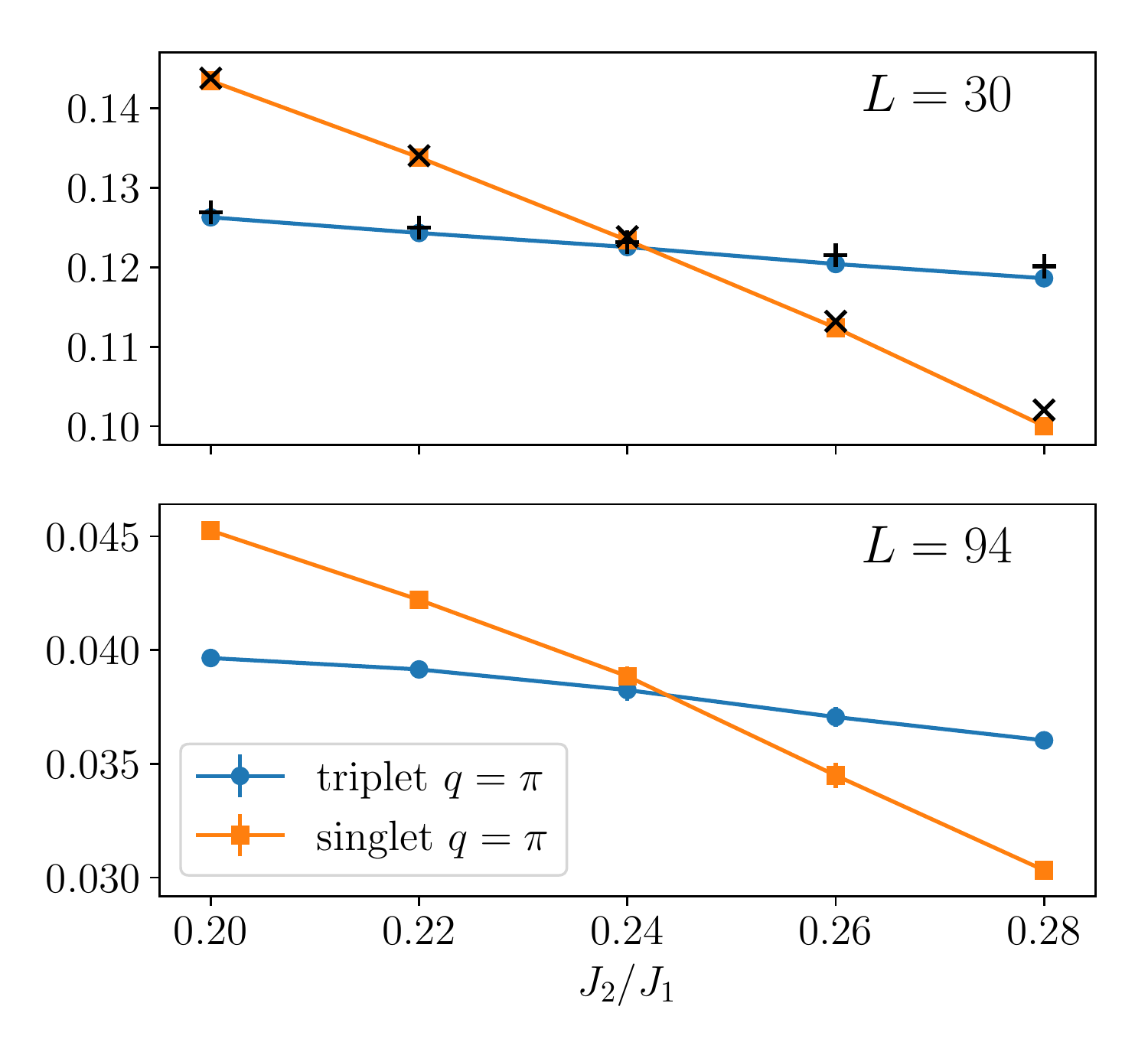}
\caption{\label{fig:cross1d} 
Triplet and singlet gaps for $L=30$ (upper panel) and $L=94$ (lower panel) for different values of the frustrating ratio $J_2/J_1$ in the 
one-dimensional $J_1-J_2$ model. The exact results for $L=30$ are also shown for comparison (black crosses).}
\end{figure}

\begin{figure}
\includegraphics[width=0.49\columnwidth]{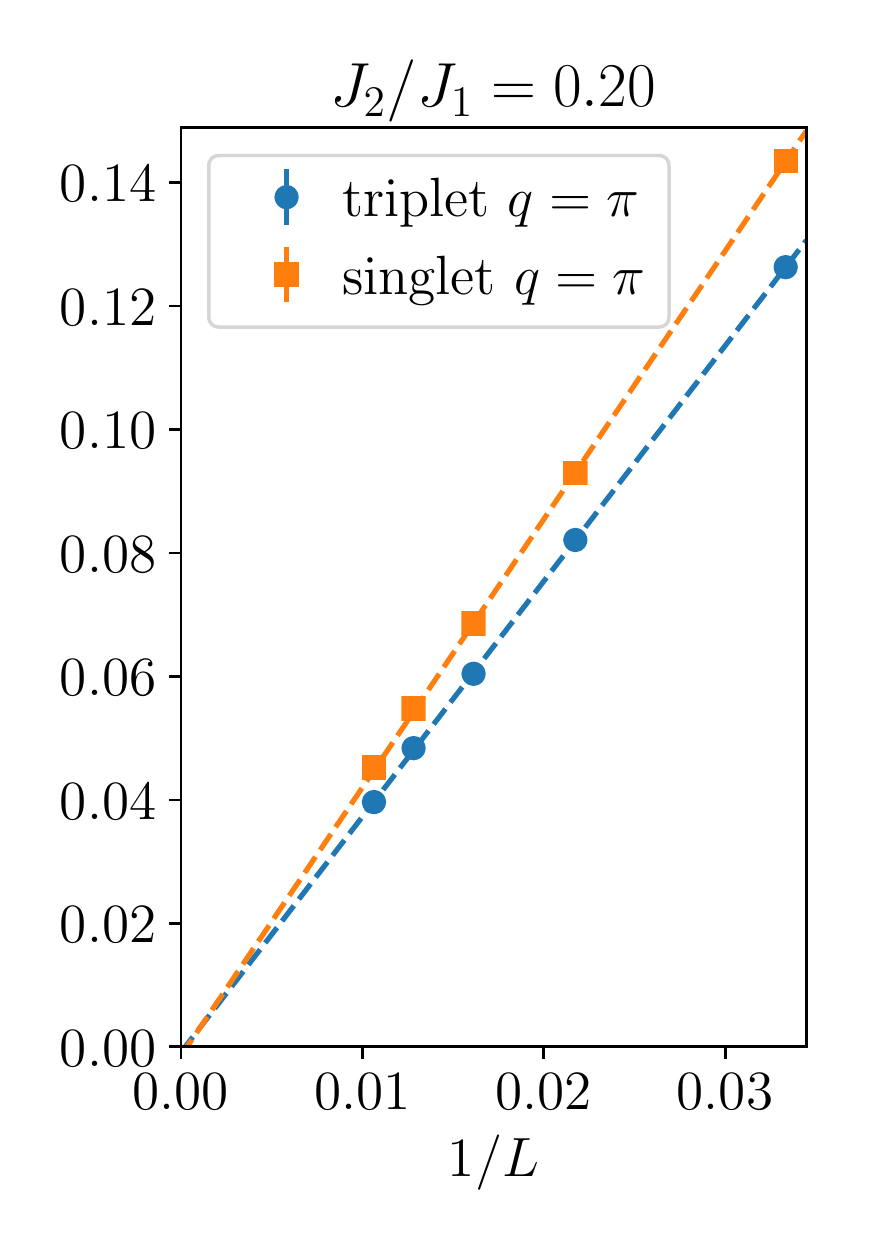}
\includegraphics[width=0.49\columnwidth]{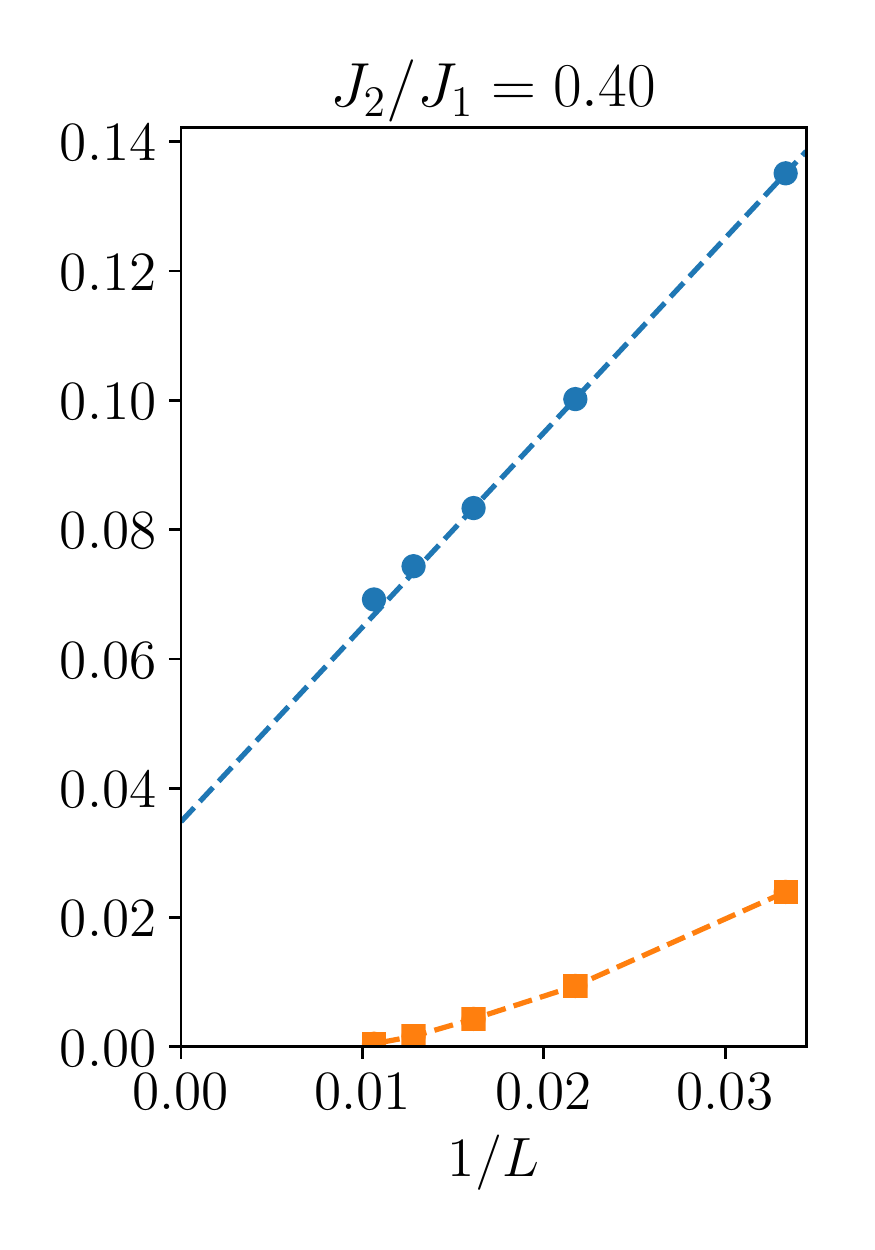}
\caption{\label{fig:scaling1d}
Size scaling of the triplet and singlet gaps of the one-dimensional $J_1-J_2$ model for two values of $J_2/J_1$.}
\end{figure}
%%%%%%%%%%%%%%%%%%%%%%%%%%%%%%%%%%%%%%%%%%%%%%%%%%%%%%%%%%%%%%%%%%%%%%%%%%%%%%%%%%%%%%%%%%%%%%%%%%%%%%%%%%%%%%

\emph{Model and methods-} 
The Hamiltonian of the $J_1-J_2$ Heisenberg model on the square lattice is defined by
\begin{equation}
\mathcal{H}=J_1 \sum_{\langle R,R^\prime \rangle} \mathbf{S}_R \cdot \mathbf{S}_{R^\prime} + 
J_2 \sum_{\langle\langle R,R^\prime \rangle\rangle} \mathbf{S}_R \cdot \mathbf{S}_{R^\prime},
\end{equation}
where $\langle \cdots \rangle$ and $\langle\langle \cdots \rangle\rangle$ indicate pairs of nearest- and next-nearest-neighboring sites, respectively.
The phase diagram of the system features two magnetic phases: for $J_2/J_1 \lesssim 0.48$ the ground state displays N\'eel order, while for 
$J_2/J_1 \gtrsim 0.6$ a different magnetic order establishes, in which the spins align ferromagnetically in one direction, and antiferromagnetically 
in the other direction. According to recent 
investigations~\cite{sushkov2001,mambrini2006,richter2010,gong2014,haghshenas2018,jiang2012,wang2013,hu2013,poilblanc2017,liu2018,hering2019,ferrari2018b}, 
the ground state between these two magnetic phases does not exhibit any magnetic order.

In this work, we focus on the non-magnetic region $0.48 \lesssim J_2/J_1 \lesssim 0.6$ and we approximate the ground state and the excited states 
of the system by means of variational wave functions based on Gutzwiller-projected fermions. To formulate our variational guess for the ground state 
of the spin system, we introduce an auxiliary BCS Hamiltonian of Abrikosov fermions,
\begin{equation}
\mathcal{H}_0=\sum_{R,R^\prime,\sigma} t_{R,R^\prime} c^\dagger_{R,\sigma} c^\dagga_{R^\prime,\sigma} + 
\sum_{R,R^\prime} \Delta_{R,R^\prime} c^\dagger_{R,\uparrow} c^\dagger_{R^\prime,\downarrow} + h.c.,
\end{equation}
and we apply the Gutzwiller projector ${\mathcal{P}_G=\prod_R n_R(2-n_R)}$ to its ground state $|\Phi_0\rangle$. The Gutzwiller projection 
suppresses all the fermionic configurations with empty or doubly occupied sites, thus returning a suitable wave function for spins, namely 
$|\Psi_0\rangle=\mathcal{P}_G|\Phi_0\rangle$. In the non-magnetic region of the $J_1-J_2$ model the best variational {\it Ansatz} for the ground 
state is obtained by a BCS Hamiltonian with a $s$-wave hopping at first-neighbors, a $d_{x^2-y^2}$-pairing at first- and fourth-neighbors, and a 
$d_{xy}$ pairing at fifth-neighbors~\cite{capriotti2001,hu2013}. The optimal values of the hopping and pairing parameters are obtained by minimizing 
the variational energy with the stochastic reconfiguration technique~\cite{sorella2005}. We note that $|\Psi_0\rangle$ is a singlet state which 
possesses all the symmetries of the lattice~\cite{wen2002}, and we emphasize that any possible attempt to break translational symmetry within the 
auxiliary BCS Hamiltonian (e.g., including a dimerized hopping or pairing) does not lead to any energy gain with respect to the uniform {\it Ansatz} 
(in contrast to the one-dimensional case~\cite{ferrari2018a}).

Following the idea of Li and Yang~\cite{li2010}, the fermionic formalism can be also employed to design variational {\it Ans\"atze} for the excited 
states of the spin model. The variational scheme is based on the definition of a set of projected particle-hole excitations:
\begin{equation} \label{eq:qR}
|q,R\rangle_{\pm} =   \mathcal{P}_G \sum_{R^\prime} e^{iq R^\prime} \left ( c^\dagger_{R+R^\prime,\uparrow}c^\dagga_{R^\prime,\uparrow}
\pm c^\dagger_{R+R^\prime,\downarrow}c^\dagga_{R^\prime,\downarrow} \right ) |\Phi_0\rangle.
\end{equation}
Here $|q,R\rangle_{+}$ and $|q,R\rangle_{-}$ are, respectively, singlet and triplet states with momentum $q$, which are labelled by the Bravais 
lattice vector $R$. Taking suitable linear combinations of these states, we can accurately approximate the low-energy singlet and triplet excitations
of the spin model. In particular, the coefficients of the linear combinations are determined by the Rayleigh-Ritz variational principle, i.e., by 
diagonalizing the Hamiltonian of the $J_1-J_2$ model within the subset $\{|q,R\rangle_{+}\}_R$, for the singlet sector, and $\{|q,R\rangle_{-}\}_R$, 
for the ${S^z=0}$ triplet sector~\cite{ferrari2018a}. The level crossing is analyzed by looking at the lowest-energy triplet and singlet wave 
functions constructed in this way.

%%%%%%%%%%%%%%%%%%%%%%%%%%%%%%%%%%%%%%%%%%%%%%%%%%%%%%%%%%%%%%%%%%%%%%%%%%%%%%%%%%%%%%%%%%%%%%%%%%%%%%%%%%%%%%
\begin{figure}
\includegraphics[width=0.9\columnwidth]{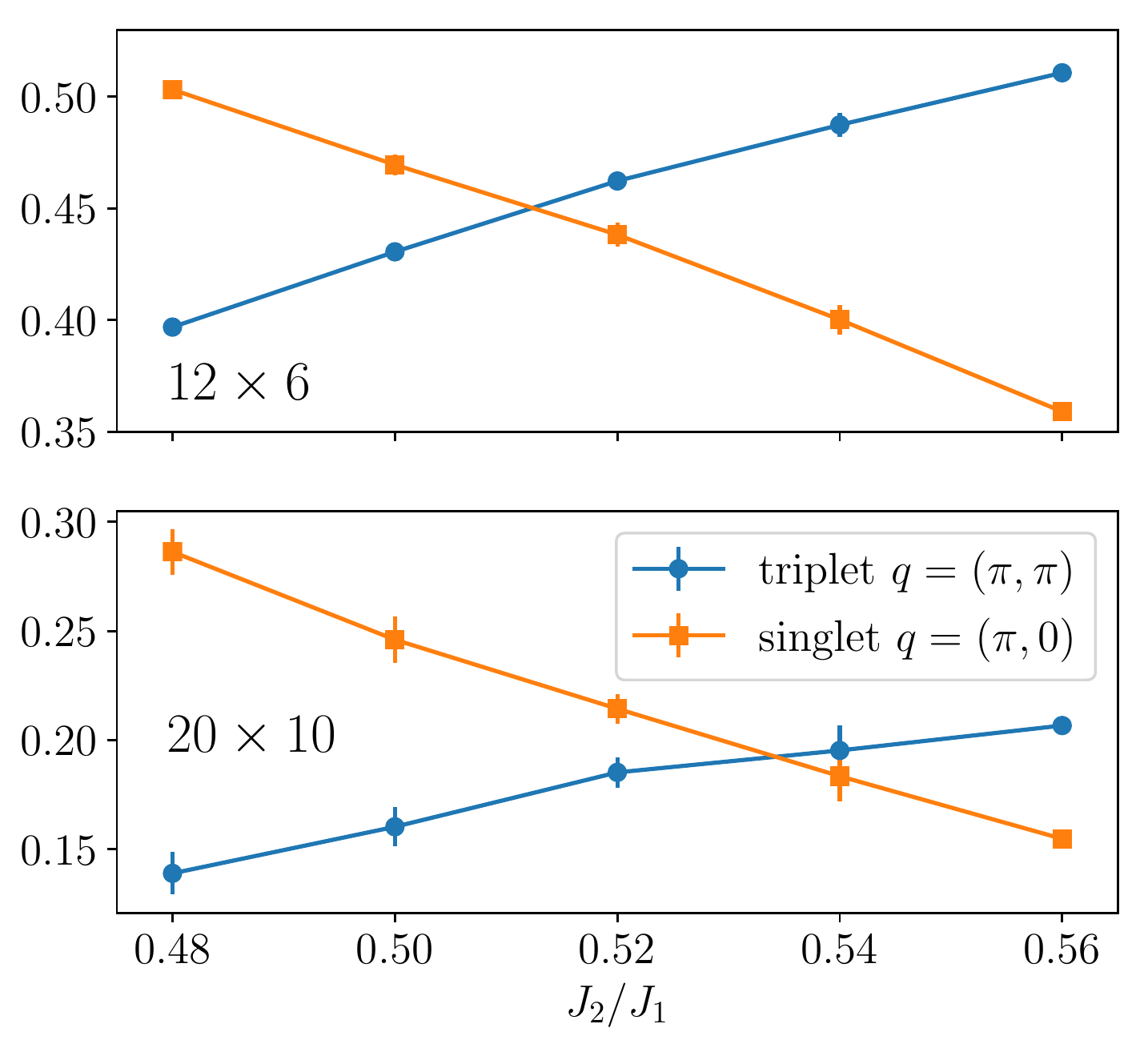}
\caption{\label{fig:crossladd} 
Triplet [with $q=(\pi,\pi)$] and singlet [with $q=(\pi,0)$] gaps for $2L \times L$ clusters with $L=6$ (upper panel) and $L=10$ (lower panel) for 
different values of the frustrating ratio $J_2/J_1$.}
\end{figure}

\begin{figure}
\includegraphics[width=0.9\columnwidth]{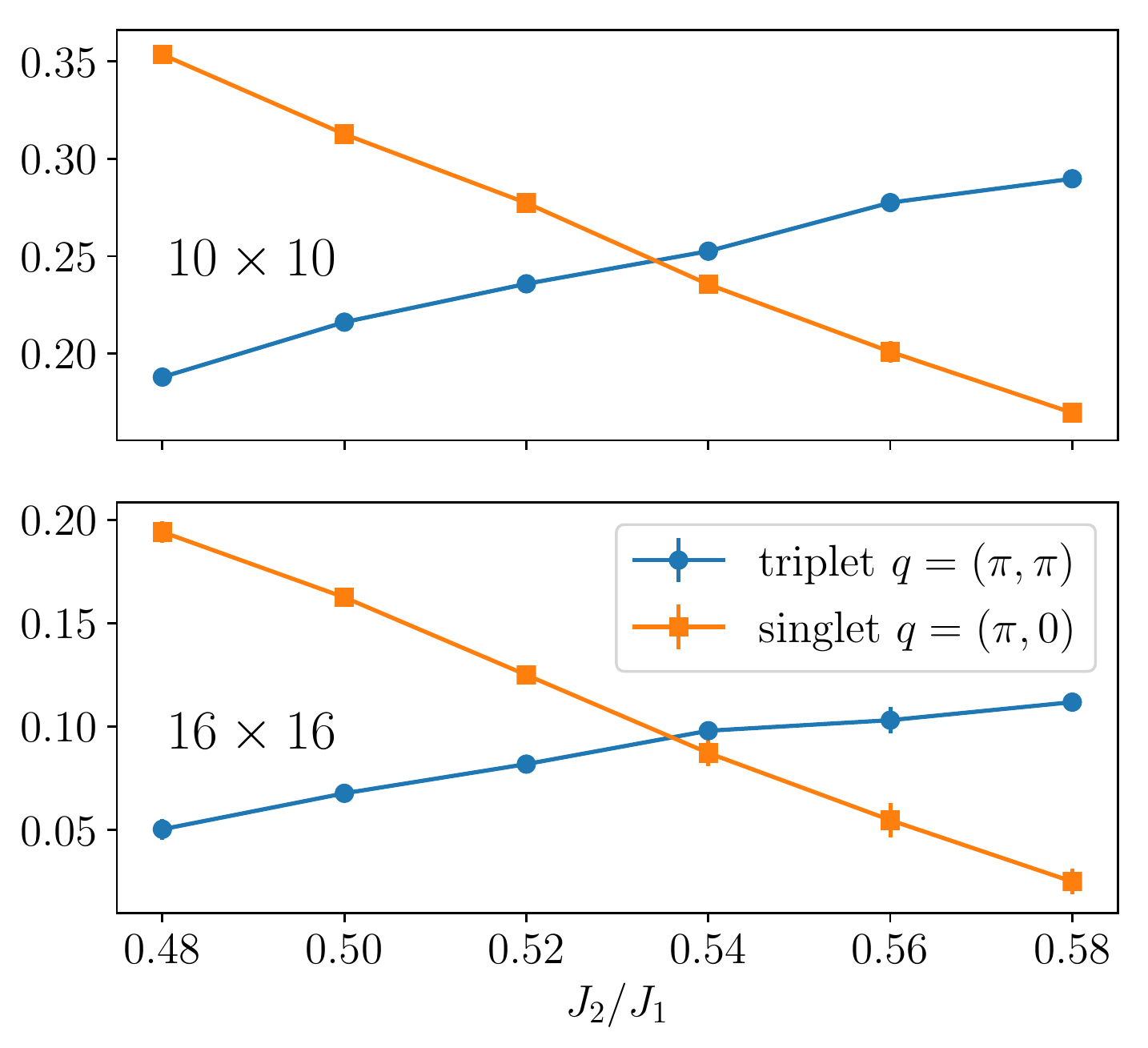}
\caption{\label{fig:cross2d} 
The same as in Fig.~\ref{fig:crossladd} for $L \times L$ clusters with $L=10$ (upper panel) and $L=16$ (lower panel).}
\end{figure}
%%%%%%%%%%%%%%%%%%%%%%%%%%%%%%%%%%%%%%%%%%%%%%%%%%%%%%%%%%%%%%%%%%%%%%%%%%%%%%%%%%%%%%%%%%%%%%%%%%%%%%%%%%%%%%

\emph{Results-} 
In order to benchmark our variational calculations, we consider the one-dimensional $J_1-J_2$ model. In this case, the transition between the gapless
phase (stable for small $J_2/J_1$) and the gapped dimerized one (stable for large $J_2/J_1$) has been located by looking at the singlet-triplet level 
crossing~\cite{okamoto1992,castilla1995}, with a very high level of accuracy of the transition point~\cite{eggert1996}, i.e., $J_2/J_1=0.241167(5)$.
In the one-dimensional case, both triplet and singlet excitations have $q=\pi$ (with respect to the ground state). The variational wave function used 
in this case is the same as the one that has been used in our previous calculations~\cite{ferrari2018a}, which is constructed from an auxiliary BCS
Hamiltonian that contains a first-neighbor hopping, plus on-site and second-neighbor pairings. In Fig.~\ref{fig:cross1d}, we report the gaps for a 
small cluster with $L=30$ (where exact diagonalizations are available) and on a large one with $L=94$. The comparison with exact results proves the 
accuracy of our estimations of the variational gaps, confirming that the phase transition appears in the correct place also when increasing the 
lattice size (the aim of this calculation is not to compete with previous estimations of the critical value). The size scaling of the gaps confirm 
gapless excitations for small frustrating ratios; by contrast, inside the dimerized phase, the triplet is gapped, while the singlet is collapsing 
to the ground state exponentially (the thermodynamic value is consistent with zero within a few errorbars). These results are shown in 
Fig.~\ref{fig:scaling1d}. In the vicinity of the transition, the triplet gap is exponentially small and, therefore, it is extremely hard to detect 
a finite value from an unbiased size scaling. In this respect, the transition is much better located by looking at the level crossing.

\begin{figure}
\includegraphics[width=0.9\columnwidth]{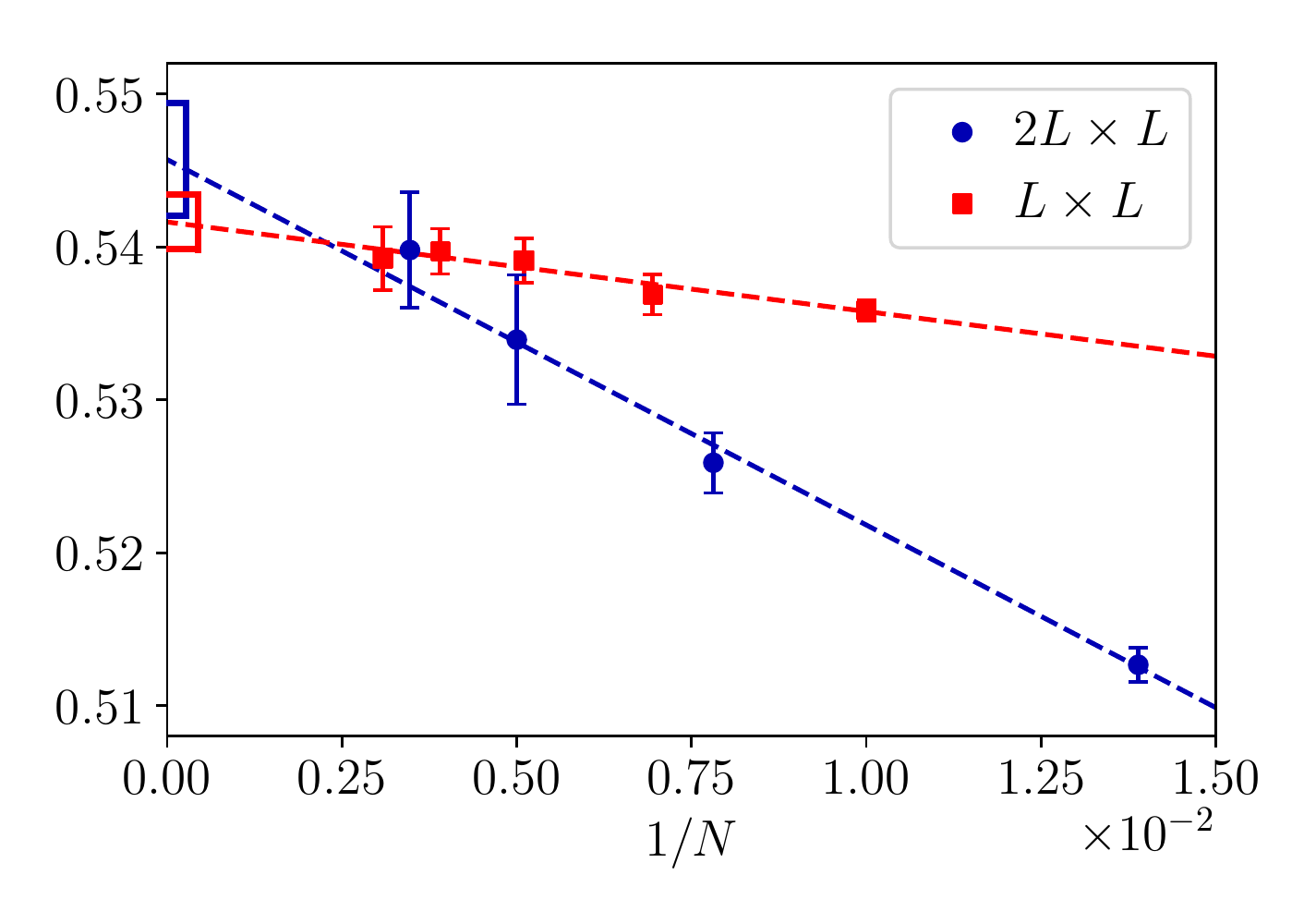}
\caption{\label{fig:scaling} 
Size-scaling of the singlet-triplet crossing point $(J_2/J_1)_c$ for ${2L \times L}$ and ${L \times L}$ geometries. Data are extrapolated as a 
function of $1/N$, where $N$ is the total number of sites (analogously to what is done in Ref.~\cite{wang2018}).}
\end{figure}

Being confident that our variational method is able to reproduce the correct features of the lowest-energy triplet and singlet excitations, we move
to the most interesting two-dimensional model. First of all, we report in Fig.~\ref{fig:crossladd}, our results for the $2L \times L$ geometry, used
in Ref.~\cite{wang2018} (the only difference is that, here, we consider periodic-boundary conditions on both directions, suitable for a 
translational-invariant wave function). Here, we find a level crossing between the triplet with $q=(\pi,\pi)$ and the singlet with $q=(\pi,0)$,
similarly to what has been obtained within density-matrix renormalization group. We would like to mention that, within this geometry, the crossing 
point moves from $J_2/J_1 \approx 0.51$ for $L=6$ to $J_2/J_1 \approx 0.535$ for $L=10$, in qualitative agreeent with Ref.~\cite{wang2018}.
Similar results can be obtained within $L \times L$ clusters, see Fig.~\ref{fig:cross2d}. The advantage of this kind of geometry, besides having 
all the point-group symmetries of the square lattice, is that the crossing point does not move substantially when changing the value of $L$. 
The size scaling of the crossing point is reported in Fig.~\ref{fig:scaling}, for both $2L \times L$ and $L \times L$ clusters. The best fit is 
obtained with the latter choice, giving $(J_2/J_1)_c=0.542(2)$; while the former one gives $(J_2/J_1)_c=0.546(4)$. 

%%%%%%%%%%%%%%%%%%%%%%%%%%%%%%%%%%%%%%%%%%%%%%%%%%%%%%%%%%%%%%%%%%%%%%%%%%%%%%%%%%%%%%%%%%%%%%%%%%%%%%%%%%%%%%

\begin{figure*}
\includegraphics[width=0.65\columnwidth]{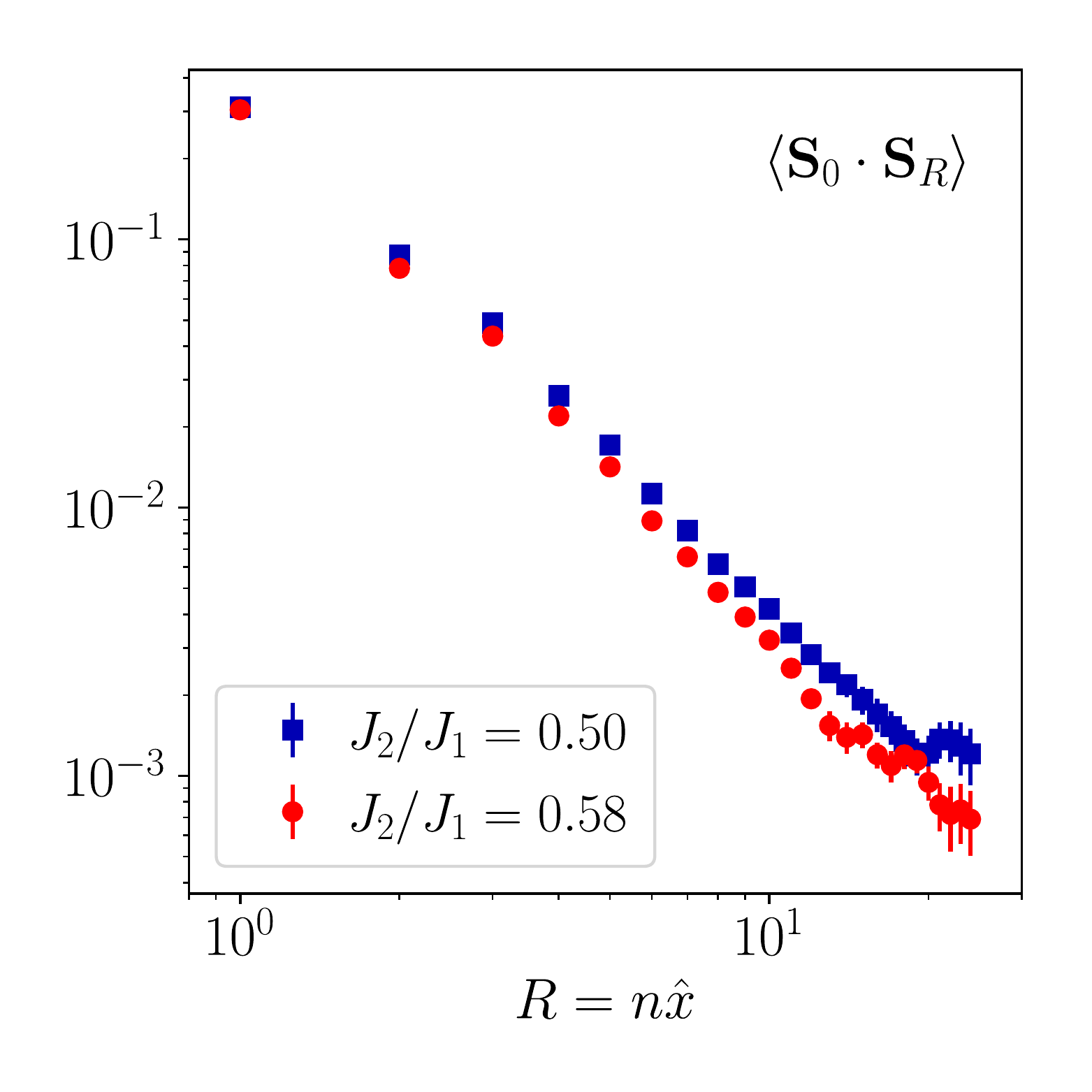}
\includegraphics[width=0.65\columnwidth]{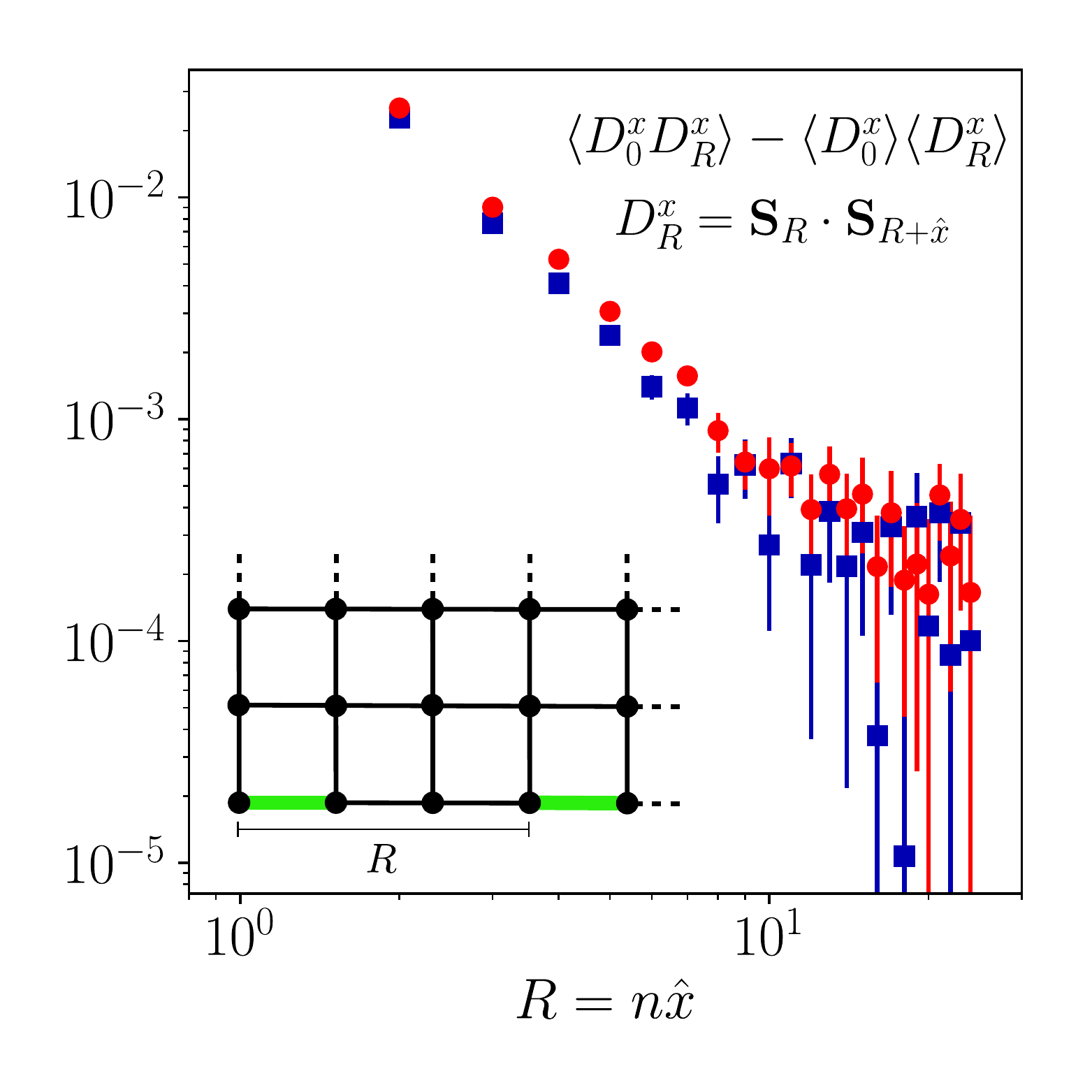}
\includegraphics[width=0.65\columnwidth]{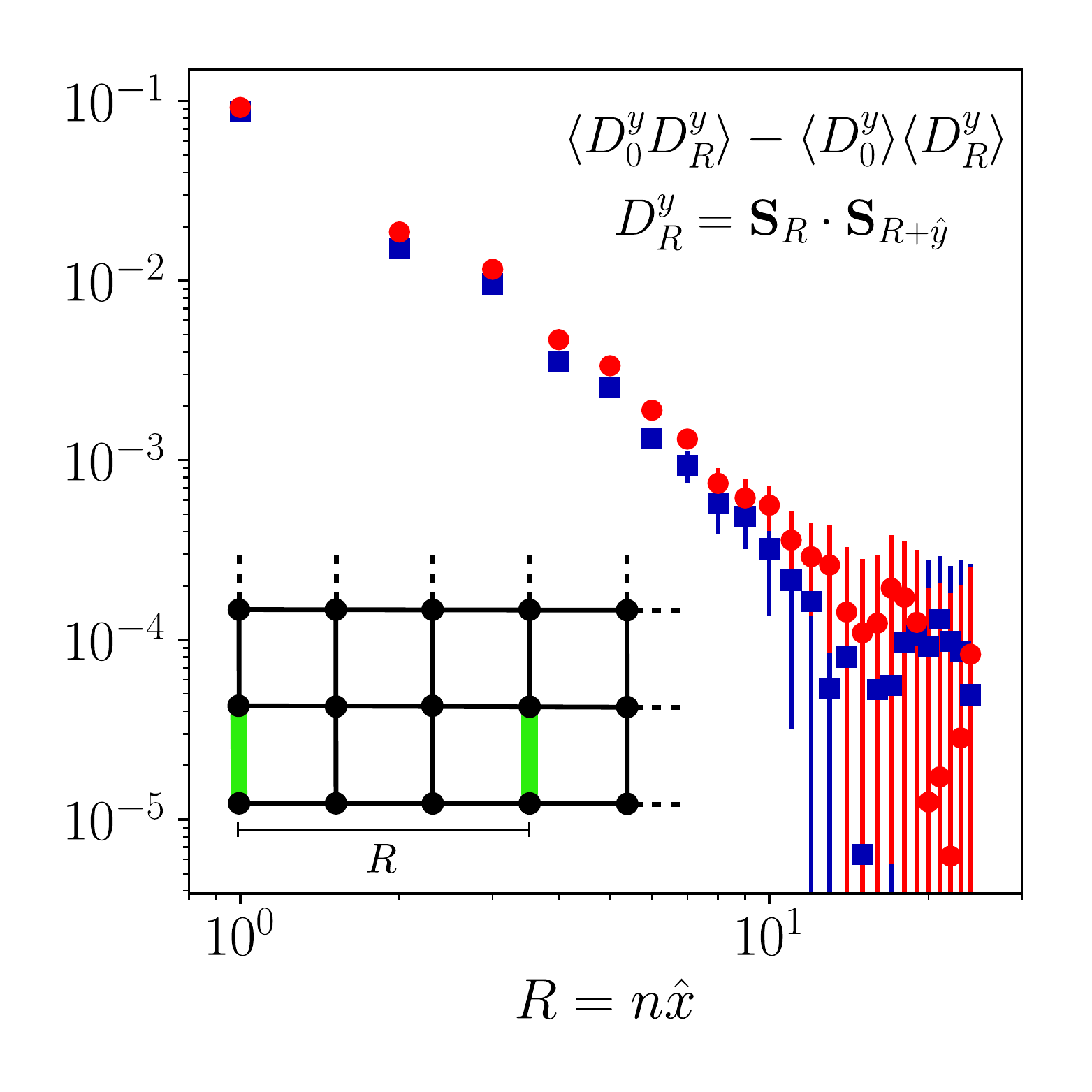}
\caption{\label{fig:dimer} 
Spin-spin and dimer-dimer correlations on the $48\times 48$ cluster, for ${J_2/J_1=0.5}$ and ${J_2/J_1=0.58}$.}
\end{figure*}
%%%%%%%%%%%%%%%%%%%%%%%%%%%%%%%%%%%%%%%%%%%%%%%%%%%%%%%%%%%%%%%%%%%%%%%%%%%%%%%%%%%%%%%%%%%%%%%%%%%%%%%%%%%%%%

Considering that the magnetic N\'eel order vanishes at $J_2/J_1 \approx 0.48$~\cite{ferrari2018b}, the present results are compatible with the 
existence of a spin-liquid region in its vicinity, i.e., for $0.48 \lesssim J_2/J_1 \lesssim 0.54$. Beyond that, it is reasonable to expect
a different phase, presumably with columnar dimer order. This conclusion is corroborated by the fact that the lowest-energy singlet excitations
have $q=(\pi,0)$ and $q=(0,\pi)$. The possibility of staggered dimers or a plaquette valence-bond order is not probable, since the ground-state 
manifold would also include a singlet at $q=(\pi,\pi)$, which instead lies much higher in energy. 

With the aim of assessing the properties of the variational state $|\Psi_0 \rangle$, we computed the isotropic spin-spin 
$\langle \mathbf{S}_{0} \cdot \mathbf{S}_{R} \rangle$ and dimer-dimer $\langle D_0^{\mu} D_{R}^\mu \rangle$ correlations 
[${D^{\mu}(R)= \mathbf{S}_{R} \cdot \mathbf{S}_{R+\hat{\mu}}}$, with $\mu=x$ or $y$]. The results are reported in Fig.~\ref{fig:dimer} for
${J_2/J_1=0.5}$ and $0.58$, which lie in the two different regions, sufficiently far away from the crossing point. Remarkably, the behavior is 
similar in both cases (at least for the largest cluster used in this work, i.e. $48 \times 48$). The spin-spin correlations have a power-law 
decay, with exponents that are consistent with $2$, namely $2.0(1)$ for $J_2/J_1=0.5$ and $2.1(1)$ for $J_2/J_1=0.58$. Also the dimer-dimer 
correlations do not show an appreciable difference between these two values of $J_2$, displaying a power-law decay in both cases (in this case,
evaluating the exponent is much harder, given the rapid decay of the signal). At first glance, these results may suggest an extremely large 
correlation length that persists in the whole region where dimer order is expected from the level-crossing analysis. Alternatively, it is 
possible that the {\it Ansatz} used for this region can be improved, including an explicit dimer order (as we discussed above, we could not
find any simple way to include a symmetry breaking that provides a variational energy gain). This hypotetical wave function may display
a much clearer evidence for dimerization, even in the correlation functions. In this respect, we want to mention that in one spatial 
dimension the level crossing can be (accurately) detected also using a simple wave function with only nearest-neighbor hopping, which has 
power-law dimer-dimer correlations (not shown). However, as discussed above, it is possible to have a substantial energy gain by including 
pairing terms in the fermionic Hamiltonian and obtain a wave function with finite dimer-dimer correlations at long distances, without affecting 
the level crossing between the lowest-energy triplet and singlet. In this regard, we definitively believe that our results on the level crossing 
in the two-dimensional case are peculiar features of the model and do not depend crucially on the details of the wave function, provided 
sufficiently accurate {\it Ans\"atze} are considered. On the other hand, we note that detecting the valence-bond ordered phase by just looking 
at the ground-state properties is a complicated task, since the dimer order is limited to a very narrow region of the phase diagram.

\emph{Conclusions-} 
By using a variational approach based on Gutzwiller-projected fermionic wave functions, we showed the existence of a level crossing between
the lowest-energy triplet, with $q=(\pi,\pi)$, and singlet, with $q=(\pi,0)$ and $(0,\pi)$, excitations within the paramagnetic region of the
$J_1-J_2$ Heisenberg model on the square lattice. Our results are in excellent agreement with recent density-matrix renormalization group
calculation by Wang and Sandvik~\cite{wang2018}, with a tiny difference in locating the N\'eel-to-spin-liquid transition ($J_2/J_1=0.48$ vs $0.46$) 
and the singlet-triplet level crossing ($J_2/J_1=0.54$ vs $0.52$). Most probably, the level crossing indicates the existence of a critical point
separating the gapless spin liquid and a valence-bond solid, with columnar order (a staggered dimer order, as well as a plaquette valence-bond order,
can be ruled out by observing that the lowest singlet with $q=(\pi,\pi)$ remains high in energy). Remarkably, within the variational {\it Ansatz} 
used in this work (on clusters up to $48 \times 48$) spin-spin correlations display a power-law behavior also beyond the level crossing; in addition, 
no visible order parameter is seen from dimer-dimer correlations. These facts may be explained by an extremely large correlation length.
Even if we could not exclude the possibility that the level crossing marks a more unconventional phase transition, we emphasize that looking
to the low-energy spectrum represents a very powerful and valuable tool to establish the ultimate phase diagram of frustrated spin models.

\emph{Acknowledgments-} 
We thank L. Wang and A. Sandvik for useful discussions. F.F. acknowledges support from the Alexander von Humboldt Foundation.

\end{document}